\documentclass[12pt]{article}
\usepackage{graphicx}
\newcommand{\bb}{\begin{equation}}
\newcommand{\ee}{\end{equation}}
\newcommand{\ba}{\begin{eqnarray}}
\newcommand{\ea}{\end{eqnarray}}

\begin{document}

\title{{\bf Inspiral Time Probability Distribution\\
for Two Black Holes Captured by\\
Emitting Gravitational Radiation}}

\author{
Don N. Page
\thanks{Internet address:
profdonpage@gmail.com}
\\
Department of Physics\\
4-183 CCIS\\
University of Alberta\\
Edmonton, Alberta T6G 2E1\\
Canada
}

\date{2024 May 21}

\maketitle
\large
\begin{abstract}
\baselineskip 20 pt

If two initially unbound black holes of masses $M_1$ and $M_2$, total mass $M = M_1 + M_2$, reduced mass $\mu = M_1 M_2/(M_1+M_2)$, and initial relative velocity $v \ll c\,(4\mu/M)$ in otherwise empty space are captured into a bound orbit by emitting gravitational radiation, the inspiral time to coalescence increases monotonically to infinity as the impact parameter $b$ approaches from below the critical impact parameter $b_c = [340\pi G^7 M^6 \mu/(3 c^5 v^9)]^{1/7} 
= [(85\pi/384)(4\mu/M)]^{1/7}(2GM/c^2)(v/c)^{-9/7}$ for capture.  Assuming a uniform flux of impinging black holes with $b \leq b_c$, the cumulative probability for impact parameters smaller than some value $b$, conditional upon the impact parameter being smaller than $b_c$, is $P = (b/b_c)^2$.  Then it is shown that the inspiral time for $[Mv^2/(4\mu c^2)]^{2/7} \ll P < 1$ is 
$T = (2\pi GM/v^3)P^{21/4}\zeta(3/2,1-P^{7/2})$, and closed-form approximate expressions for the inverse function $P(T/T_0)$ with $T_0 = 2\pi GM/v^3$ are also given.

\end{abstract}

\normalsize

\baselineskip 22 pt

\newpage

\section{Introduction}

LIGO and Virgo have had enormous success in detecting the gravitational waves from astrophysical inspiraling black holes that coalesce to form a larger black hole
\cite{
LIGOScientific:2016aoc,
LIGOScientific:2016emj,
LIGOScientific:2016vbw,
LIGOScientific:2016vlm,
LIGOScientific:2016lio,
LIGOScientific:2016kwr,
LIGOScientific:2016vpg,
LIGOScientific:2016sjg,
LIGOScientific:2016dsl,
LIGOScientific:2017bnn,
LIGOScientific:2017ycc,
LIGOScientific:2017vox,
LIGOScientific:2018dkp,
LIGOScientific:2018mvr,
LIGOScientific:2018jsj,
LIGOScientific:2019lzm,
LIGOScientific:2019fpa,
LIGOScientific:2020aai,
LIGOScientific:2020stg,
LIGOScientific:2020zkf,
LIGOScientific:2020iuh,
LIGOScientific:2020ufj,
LIGOScientific:2020ibl,
LIGOScientific:2020tif,
LIGOScientific:2021usb,
LIGOScientific:2021djp,
LIGOScientific:2021sio,
Ghosh:2022xhn}.
Of course LIGO and Virgo can only observe the final stages of the inspiral and coalescence.  It is therefore of interest to try to estimate initial conditions for the inspiral and the subsequent evolution.  For example, it would be desirable to have estimates for the eccentricity of the inspiral orbits as they enter the LIGO and Virgo observational regimes.

Here an attempt at a partial answer is given under the simplified assumption that before they become gravitationally bound, the two black holes, of masses $M_1$ and $M_2$, are approaching each other at relative velocity $v \ll c$ and with a uniform probability distribution in the transverse plane (probability per small range of impact parameter proportional to the impact parameter).  There is then a critical impact parameter $b_c$, proportional to the Schwarzschild radius of the total mass $M = M_1 + M_2$, multiplied by the 7th root of the ratio of the reduced mass $\mu = M_1 M_2/M$ to the total mass $M$, and divided by the 9/7 power of the ratio of the initial velocity to the speed of light, $v/c$, such that for impact parameters $b$ larger than $b_c$, the two black holes scatter without becoming bound, but for $b < b_c$, the black holes emit enough radiation in their initial close encounter that they become bound.  Conditionalizing on $b < b_c$, the cumulative probability for an impact parameter smaller than $b$ is then 
\bb
P = \left(\frac{b}{b_c}\right)^2,
\label{P}
\ee 
the fraction $(b/b_c)^2$ of the total area inside the capture disk of radius $b_c$ that also has the impact parameter vector of magnitude smaller than $b$.  

\section{Calculating the Critical Impact Parameter $b_c$}

I am assuming two black holes of individual masses $M_1$ and $M_2$, total mass $M = M_1 + M_2$, reduced mass $\mu = M_1 M_2/M$, and initial relative velocity $v \ll c$ in the center-of-momentum (COM) frame in which the initial velocity directions are such that the straight lines through them would pass at an impact parameter $b$ that is much larger than the Schwarzschild radius $r_S = 2GM/c^2$ corresponding to the total mass $M$.  In the COM, the initial energy (not counting the rest mass energies) is $E_0 = (1/2)\mu v^2$, and the initial angular momentum is $L_0 = \mu v b$.

Calculations by Philip Carl Peters and by Peters and Jon Mathew
\cite{Peters:1963ux,Peters:1964qza,Peters:1964zz} show that for a nonrelativistic Keplerian orbit of semimajor axis $a$ and eccentricity $e$, the power radiated into gravitational waves, averaged over one orbit, is (Eq.\ (16) in \cite{Peters:1963ux}, the same copied on page 92 of \cite{Peters:1964qza}, and Eq.\ (5.4) of \cite{Peters:1964zz})
\ba
\left\langle -\frac{dE}{dt}\right\rangle &=&
\frac{32}{5}\frac{G^4}{c^5}\frac{M_1^2 M_2^2(M_1+M_2)}{a^5(1-e^2)^{7/2}}
\left(1 + \frac{73}{24}e^2 + \frac{37}{96}e^4\right) \nonumber \\
&=& \frac{32}{5}\frac{G^4}{c^5}\frac{M^3\mu^2}{a^5(1-e^2)^{7/2}}
\left(1 + \frac{73}{24}e^2 + \frac{37}{96}e^4\right).
\label{power}
\ea
Similarly, the angular momentum radiated into gravitational waves, averaged over one orbit, is (Eq.\ (5.32) on page 99 of \cite{Peters:1964qza}, and Eq.\ (5.5) of \cite{Peters:1964zz})
\ba
\left\langle -\frac{dL}{dt}\right\rangle &=&
\frac{32}{5}\frac{G^{7/2}}{c^5}\frac{M_1^2 M_2^2(M_1+M_2)^{1/2}}
{a^{7/2}(1-e^2)^2}
\left(1 + \frac{7}{8}e^2\right) \nonumber \\
&=& \frac{32}{5}\frac{G^{7/2}}{c^5}\frac{M^{5/2}\mu^2}{a^{7/2}(1-e^2)^2}
\left(1 + \frac{7}{8}e^2\right).
\label{torque}
\ea

Since a Keplerian elliptical orbit has energy $E = -GM_1 M_2/(2a) = -GM\mu/(2a)$, angular momentum $L = \sqrt{GM\mu^2 a(1-e^2)}$, and orbital period $\tau = 2\pi\sqrt{a^3/(GM)}$, the energy and angular momentum radiated during one orbit are
\ba
-\Delta E &=& \tau\left\langle -\frac{dE}{dt}\right\rangle 
 = \frac{64\pi}{5}\frac{G^{7/2}M^{5/2}\mu^2}{c^5 a^{7/2}(1-e^2)^{7/2}}
\left(1 + \frac{73}{24}e^2 + \frac{37}{96}e^4\right),
\label{DeltaE}\\
-\Delta L &=& \tau\left\langle -\frac{dL}{dt}\right\rangle
 = \frac{64\pi}{5}\frac{G^3 M^2 \mu^2}{c^5 a^2 (1-e^2)^2}
\left(1 + \frac{7}{8}e^2\right).
\label{DeltaL}
\ea
Then making the substitutions $a(1-e^2) = L^2/(GM\mu^2)$ and
\bb
e^2 = 1 - \frac{-2EL^2}{G^2M^2\mu^3}
\label{e2}
\ee
gives the dimensionless ratios
\bb
-\frac{\Delta E}{\mu c^2} = \frac{170\pi}{3}\left(\frac{GM\mu}{cL}\right)^7
\frac{\mu}{M}\left[1-\frac{366}{425}\left(\frac{-2EL^2}{G^2M^2\mu^3}\right)
+\frac{37}{425}\left(\frac{-2EL^2}{G^2M^2\mu^3}\right)^2\right],
\label{DeltaE/restmass}
\ee
\bb
-\frac{\Delta L}{L} = 24\pi\left(\frac{GM\mu}{cL}\right)^5 \frac{\mu}{M}
\left[1-\frac{7}{15}\left(\frac{-2EL^2}{G^2M^2\mu^3}\right)\right].
\label{Deltal/l}
\ee

For two black holes of total mass $M = M_1+M_2$ and reduced mass $\mu = M_1M_2/M$ approaching each other from far away with initial incoming relative velocity $v \ll c$ and impact parameter $b$, one has initial energy $E_0 = (1/2)\mu v^2$ and angular momentum $L_0 = \mu b v$ in the COM frame.  I shall assume that the impact parameter $b$ obeys the following inequalities,
\bb
\frac{2GM}{c^2}\left(\frac{c}{v}\right)
= \frac{2GM}{cv} \ll b \ll \frac{2GM}{v^2}
= \frac{2GM}{c^2}\left(\frac{c}{v}\right)^2,
\label{binequalities}
\ee
the inequality for the lower bound of $b$ so that the distance of closest approach on the first pass is much greater than the Schwarzschild radius $2GM/c^2$ corresponding to the total mass $M$ (so that the nonrelativistic equations of Peters and Mathews \cite{Peters:1963ux,Peters:1964qza,Peters:1964zz} for the gravitational radiation apply), and the inequality for the upper bound of $b$, which implies that the eccentricity $e$ is very near 1, so that the deflection angle is nearly $\pi$ radians (which is a necessary but not sufficient condition for the initially unbound black holes to emit enough gravitational radius to become bound, with $E$ becoming negative, of course not counting the rest mass energy $Mc^2$ in these nonrelativistic calculations of the orbits).

The condition that the two black holes, originally in a hyperbolic but nearly parabolic orbit unbound with positive energy $E_0 = (1/2)\mu v^2$ and angular momentum $L_0 = \mu b v$, lose enough energy to gravitational radiation to become bound into an elliptical orbit, with negative energy $E_1 = E_0 + \Delta E < 0$, semimajor axis $a_1 = GM\mu/(-2E_1)$, and period $\tau_1 = 2\pi\sqrt{a_1^2/(GM)} = 2\pi GM\sqrt{\mu^3/(-E_1)^3}$, is that
\bb
-\Delta E = \frac{170\pi}{3}\left(\frac{GM}{cvb}\right)^7
\frac{\mu^2 c^2}{M}\left[1+\frac{366}{425}\left(\frac{bv^2}{GM}\right)^2
+\frac{37}{425}\left(\frac{bv^2}{GM}\right)^4\right]
> E_0 = \frac{1}{2}\mu v^2.
\label{DeltaE-inequality1}
\ee

This condition for capture is by itself sufficient to imply that $bv^2/(GM) \ll 1$, so the condition for capture becomes essentially the same as
\bb
-\frac{\Delta E}{E_0} =  \frac{340\pi}{3}\left(\frac{GM}{cvb}\right)^7
\frac{\mu}{M}\left(\frac{c}{v}\right)^2 > 1.
\label{DeltaE-inequality2}
\ee
This in turn implies that for capture, the impact parameter $b$ must be less than the critical impact parameter $b_c$ given by
\bb
b_c = \left(\frac{85\pi}{384}\right)^{1/7}\left(\frac{4\mu}{M}\right)^{1/7}
\left(\frac{c}{v}\right)^{9/7}\frac{2GM}{c^2} \approx
0.949\,429\left(\frac{4\mu}{M}\right)^{1/7}
\left(\frac{c}{v}\right)^{9/7}\frac{2GM}{c^2}.
\label{bc}
\ee
Here I have used $4\mu/M = 4 M_1M_2/(M_1+M_2)^2$ with the factor of 4 so that its upper bound is 1, and I have included the 2 in $2GM/c^2$ so that it is the Schwarzschild radius of the total mass $M$, which coincidentally makes the leading numerical factor only about 5\% less than of unity.

Since $4\mu/M \leq 1$ and since I have assumed $v/c \ll 1$, it is indeed the case that the second (right-hand) inequality in (\ref{binequalities}) is always satisfied for impact parameters $b$ that lead to capture, $b < b_c \ll (2GM/c^2)(c/v)^2$, so that the initial unbound orbit has eccentricity $e$ very near 1, and hence, for the initial capture at least, we can ignore the two terms in Eq.\ (\ref{DeltaE-inequality1}) with positive powers of $e^2-1 = [2EL^2/(G^2M^2\mu^3)] = [bv^2/(GM)]^2$.  On the other hand, there is not necessarily a capture impact parameter $b < b_c$ such that the first inequality in (\ref{binequalities}) is satisfied, since this would require $(2GM/c^2)(c/v) \ll b_c$, which is equivalent to $(v/c)^2 \ll 4\mu/M$, which even for $v/c \ll 1$ may not be satisfied for highly unequal black hole masses that hence have very small $4\mu/M$.  However, I shall assume $(v/c)^2 \ll 4\mu/M$ so that $(2GM/c^2)(c/v) \ll b_c$.

Furthermore, even when $(v/c)^2 \ll 4\mu/M$ so that $(2GM/c^2)(c/v) \ll b_c$ is true, for sufficiently smaller impact parameters (that is, for $b$ not much larger than $(2GM/c^2)(c/v)$), the point of closest approach during the initial unbound but nearly parabolic orbit may not be much larger than the Schwarzschild radius $2GM/c^2$, in which case the nonrelativistic formulas of Peters and Mathews do not apply.  Indeed, for highly unequal black hole masses so that $4\mu/M \ll 1$, impact parameters smaller than approximately $(4GM/c^2)(c/v) = 4GM/(cv)$ lead to direct coalescence of the black holes, without requiring first the emission of enough gravitational wave energy to lead to bound orbits before capture.  Therefore, I shall restrict my nonrelativistic analysis to initial unbound orbits with impact parameters $b$ and critical impact parameters $b_c$ in the range
\bb
\left(\frac{c}{v}\right)\frac{2GM}{c^2} = \frac{2GM}{cv} \ll b < b_c
= \left(\frac{85\pi}{384}\right)^{1/7}\left(\frac{4\mu}{M}\right)^{1/7} 
\left(\frac{c}{v}\right)^{9/7}\frac{2GM}{c^2},
\label{binequalities2}
\ee
which requires $b_c \gg 2GM/(cv)$, which in turn is equivalent to
\bb
\left(\frac{v}{c}\right)^2 \ll \frac{4\mu}{M} \leq 1.
\label{vmuinequalities}
\ee

\section{Properties of the Initial Set of Orbits}

Suppose we imagine a uniform flux of black holes at velocity $v \ll c\,$ relative to another hole, and condition on those black hole pairs whose impact parameters $b$ are less than the critical impact parameter $b_c$ given by Eq.\ (\ref{bc}).  Because $-\Delta E/E_0 = (b_c/b)^7$, the conditional probability, given by Eq.\ (\ref{P}), that a pair will have impact parameter less than some $b \leq b_c$ is
\bb
P = \left(\frac{b}{b_c}\right)^2 = \left(\frac{-\Delta E}{E_0}\right)^{-2/7}
\label{PE}
\ee

Now we would like to calculate the approximate inspiral time, from the moment of closest approach of the initially unbound orbit that becomes bound, until the coalescence of the two black holes.  We shall use only the nonrelativistic formulas of Peters and Mathews, so the error in the approximation will be at least of the order of the final coalescence time when the situation becomes highly nonlinear, which one might expect to be a few times the light travel time across a distance equivalent to the Schwarzschild radius for the total mass, a time of the order of $2GM/c^3$.

The lower bound on the impact parameter from the left inequality of (\ref{binequalities}) or (\ref{binequalities2}) is equivalent to the lower bound on the cumulative conditional probability $P = (b/b_c)^2$ given by
\bb
\left(\frac{Mv^2}{4\mu c^2}\right)^{2/7} \ll P = \left(\frac{b}{b_c}\right)^2
= \left(\frac{E_0}{-\Delta E}\right)^{2/7} 
\label{Pmin}
\ee
This inequality implies that for the nonrelativistic formulas of Peters and Mathew to apply, the loss of energy per orbit for the initial encounter, which these formulas give as $-\Delta E = E_0 P^{-7/2} = (1/2)\mu v^2 P^{-7/2}$, must be much less than $(2\mu/M)\mu c^2$, which for equal rest masses is one-eighth the total rest mass energy, requiring $-\Delta E \ll (1/8)Mc^2$, and which when $M_1/M_2 \ll 1$ is approximately twice this mass ratio multiplied by the rest mass energy of the first mass, $-\Delta E \ll 2(M_1/M_2)M_1 c^2$.

Similarly, Eq.\ (\ref{Deltal/l}) with $L$ evaluated at the initial value of the angular momentum, $L_0 = \mu b v$, and at the initial value of the energy, $E_0 = (1/2)\mu v^2$, gives
\bb
-\frac{\Delta L}{L_0} = 24\pi\left(\frac{GM}{cvb}\right)^5 \frac{\mu}{M}
\left[1+\frac{7}{15}\left(\frac{bv^2}{GM}\right)^2\right].
\label{DeltaL/L}
\ee
Again the inequality $b < b_c$ in (\ref{binequalities2}) implies that
\bb
\frac{bv^2}{GM} < \left(\frac{85\pi}{3}\right)^{1/7}
\left(\frac{4\mu}{M}\right)^{1/7}\left(\frac{v}{c}\right)^{5/7},
\label{einequality}
\ee
which is very small for $v\ll c$, so, at least initially after the black holes become captured into a bound orbit, the eccentricity is very nearly one, and we can ignore the second term inside the square brackets of Eq. (\ref{DeltaL/L}).  Then the relative change in the angular momentum per orbit initially is
\bb
-\frac{\Delta L}{L_0} \approx \frac{3\pi}{4}\left(\frac{2GM}{cvb}\right)^5 \frac{\mu}{M}.
\label{DeltaL/L2}
\ee
The left inequality of (\ref{binequalities}) or (\ref{binequalities2}) gives the further implication that the fractional decrease in the angular momentum $L$ per orbit is initially much less than 1,
\bb
-\frac{\Delta L}{L_0} \ll 1.
\label{DeltaL/Linequality}
\ee

Thus for many orbits after the two black holes become bound, their angular momentum remains nearly constant.  Furthermore, until $1-e^2 = -2EL^2/(G^2M^2\mu^3)$ (which starts out negative but very nearly zero when the two holes are unbound but then becomes positive but still very nearly zero for many orbits) becomes significantly positive (which would reduce the energy loss per orbit, as would a decrease of $L$), the energy lost per orbit given by Eq.\ (\ref{DeltaE}) or (\ref{DeltaE-inequality1}) would also remain nearly constant.  Therefore, for at least many initial orbits, the energy loss per orbit would be nearly constant, so that the energy after $n$ orbits would be approximately
\ba
E_n &\approx& E_0 + n \Delta E = \frac{1}{2}\mu v^2 
- \left[\frac{170\pi}{3}\left(\frac{GM}{cvb}\right)^7\frac{\mu^2 c^2}{M}\right]n \nonumber \\
&=& \frac{1}{2}\mu v^2 \left[1-\left(\frac{b_c}{b}\right)^7 n\right]
= \frac{1}{2}\mu v^2 \left[1-P^{-7/2}\, n\right]. 
\label{En}
\ea

Indeed, under the assumptions that the orbital angular momentum $L$ stays near $L_0 = \mu b v$ (with initial impact parameter $b$ and initial relative velocity $v$), and that also $1-e^2$ stays near zero (which we shall immediately see is a nearly equivalent condition), then one can calculate that after $n \gg 1$ orbits,
\ba
-\frac{\Delta L}{L_0} &\approx& 
\left(\frac{2^{11}3^{12}}{85^5}\pi^2\right)^{1/7}
\left(\frac{\mu}{M}\right)^{2/7}
\left(\frac{v}{c}\right)^{10/7} P^{-5/2}\, n \nonumber \\
&\approx& 1.135
\left(\frac{\mu}{M}\right)^{2/7}
\left(\frac{v}{c}\right)^{10/7} P^{-5/2}\, n,
\label{DeltaL/L3}
\ea
\ba
1 - e^2 &\approx&
\left(\frac{340\pi}{3}\right)^{2/7}
\left(\frac{\mu}{M}\right)^{2/7}
\left(\frac{v}{c}\right)^{10/7} P^{-5/2}\, n \nonumber \\
&\approx& \frac{85}{18}\left(-\frac{\Delta L}{L_0}\right) \nonumber \\
&\approx& 5.358
\left(\frac{\mu}{M}\right)^{2/7}
\left(\frac{v}{c}\right)^{10/7} P^{-5/2}\, n.
\label{1-e2}
\ea

Therefore, Eq.\ (\ref{En}) applies for
\bb
n \ll 0.2 \left(\frac{M}{\mu}\right)^{2/7}
\left(\frac{c}{v}\right)^{10/7} P^{5/2}.
\label{nineq}
\ee
In order for there to be a large number $n$ of orbits during which $L \approx L_0 = \mu b v$ and $1 - e^2 \ll 1$, one must have the conditional probability $P = (b/b_c)^2$ that the impact parameter is smaller than $b$, given that it is smaller than the critical impact parameter $b_c$ for the two black holes to become bound by the emission of gravitational radiation, obey the following inequality:
\bb
P = \left(\frac{b}{b_c}\right)^2 \gg 
\left(\frac{340\pi}{3}\frac{\mu}{M}\frac{v^5}{c^5}\right)^{4/35}
\sim 2\left(\frac{\mu}{M}\right)^{4/35}\left(\frac{v}{c}\right)^{4/7}.
\label{Pineq}
\ee
This is equivalent to having the initial impact parameter
\bb
b \gg \left(\frac{85\pi}{96}\right)^{1/5}
\left(\frac{4\mu}{M}\right)^{1/5}\frac{2GM}{cv}
\approx 1.227\left(\frac{4\mu}{M}\right)^{1/5}\frac{2GM}{cv}.
\label{bineq}
\ee
This is implied by the left inequality of (\ref{binequalities}) or (\ref{binequalities2}), $2GM/(cv) \ll b$, which also implies that
\bb
P = \left(\frac{b}{b_c}\right)^2 \gg \left(\frac{Mv^2}{4\mu c^2}\right)^{2/7}.
\label{Pstrongerineq}
\ee
Since this is a stronger inequality than Eq.\ (\ref{Pineq}), assuming the inequalities given in (\ref{binequalities}), there is always a large range of $n$ for which $L \approx L_0 = \mu b v$ and $1-e^2 \ll 1$ so that Eq.\ (\ref{En}) gives an accurate estimate of the energy $E_n$ of the $n$th orbit after the initial encounter that emits enough gravitational radiation to leave the two black holes in bound orbits.

After I independently derived Eq.\ (\ref{bc}) above for the critical impact parameter $b_c$ from the equations of Peters and Mathews \cite{Peters:1963ux,Peters:1964qza,Peters:1964zz}, an editor pointed out the paper of R.\ O.\ Hansen \cite{Hansen:1972jt}, which gave two complicated formulas relating the initial energy, initial eccentricity, and initial impact parameter for the critical cases (with fixed masses), but it did not use the fact that one can neglect $f = 1 - e^2$ to get the explicit formula of Eq.\ (\ref{bc}) above for the critical impact parameter in terms of the initial relative velocity $v$ (and hence initial energy $E_0 = (1/2)\mu v^2$, for fixed given masses).  However, I found that essentially Eq.\ (\ref{bc}) above was given by Gerald D.\ Quinlan and Stuart L.\ Shapiro \cite{Quinlan:1989}.  Hideaki Mouri and Yoshiaki Taniguchi \cite{Mouri2002} also got the same formula but from the starting point of calculations of gravitational radiation from parabolic orbits by Michael Turner \cite{Turner1977} rather than from Peters and Mathews \cite{Peters:1963ux,Peters:1964qza,Peters:1964zz}.   A later paper giving the formula (presumably independently, since not mentioning that Quinlan and Shapiro had given it, though they were cited for other reasons) is \cite{OLeary:2008myb}.  Nevertheless, my use of these results in the next section for the inspiral merger time appear to be new.

\section{Inspiral Time $T$ as a Function of the Cumulative Conditional Probability $P$}

Therefore, assuming the inequalities $2GM/(cv) \ll b < b_c \ll 2GM/v^2$ of (\ref{binequalities}) and (\ref{binequalities2}) for the initial impact parameter $b$ and critical impact parameter $b_c$ given by Eq.\ (\ref{bc}) in terms of the black hole total rest mass $M$ and reduced mass $\mu$, and with the two black holes having initial relative velocity $v$, we have Eq.\ (\ref{En}), $E_n = \frac{1}{2}\mu v^2 (1-P^{-7/2}\, n)$, applying for $1 \leq n \ll 0.2 (M/\mu)^{2/7}(c/v)^{10/7} P^{5/2}$.  Here $P = (b/b_c)^2$ is the conditional probability, assuming a uniform flux of impinging black holes, that the impact parameter is less than $b$, conditional upon its being less than $b_c$ as is necessary for the two black holes to become gravitationally bound to each other.

With these assumptions, we have that the semimajor axis of the $n$th orbit is
(remembering that $v$ is the initial relative velocity and not the orbital velocity)
\bb
a_n = \frac{GM\mu}{-2E_n} = \frac{GM}{v^2(P^{-7/2}\, n - 1)},
\label{an}
\ee
and the period of the $n$th orbit (from the $n$th close encounter to the $(n+1)$th close encounter) is
\bb
\tau_n = 2\pi\sqrt{\frac{a_n^3}{GM}} 
= \frac{2\pi GM}{v^3}\left(P^{-7/2}\, n - 1\right)^{-3/2}
= \frac{2\pi GM}{v^3}P^{21/4}\left(n - P^{7/2}\right)^{-3/2}.
\label{taun}
\ee

Using the Hurwitz zeta function,
\bb
\zeta(s,a) = \sum_{n=0}^\infty \frac{1}{(n+a)^s},
\label{HZ}
\ee
we see that the total inspiral time is, ignoring small relative corrections for\\
$n \stackrel{>}{\sim} 0.2(M/\mu)^{2/7}(c/v)^{10/7}P^{5/2}$ that violate the inequality (\ref{nineq}),
\bb
T = \frac{2\pi GM}{v^3} P^{21/4}\, \zeta\left(\frac{3}{2},1\!-\! P^{7/2}\right)
  = T_0(M,v)\, t(P),
\label{T}
\ee
a function of the total mass $M = M_1 + M_2$, the initial velocity $v$ of approach before becoming gravitationally bound, and the cumulative conditional probability $P = (b/b_c)^2$ for impact parameters smaller than $b$, given the condition that $b$ is smaller than the critical impact parameter 
$[(85\pi/384)(4\mu/M)]^{1/7}(2GM/c^2)(v/c)^{-9/7}$, 
so that for all impact parameters smaller than $b$, the two initially unbound black holes get captured into a bound orbit by the emission of gravitational radiation, except that for $b \stackrel{<}{\sim} 4GM/(cv)$, the two black holes will coalesce without orbiting, so my analysis excludes such small values of the impact parameter $b$.  The overall scale of the time, given by the total mass and initial velocity, is
\bb
T_0(M,v) = \frac{2\pi GM}{v^3} 
= (0.264\ \mathrm{year})\left(\frac{M}{10M_\odot}\right)
\left(\frac{100\ \mathrm{km/s}}{v}\right)^3,
\label{T0}
\ee
and the dimensionless factor depending on the conditional probability $P$ is
\bb
t(P) = P^{21/4}\,\zeta\left(\frac{3}{2},1\!-\! P^{7/2}\right) = \left(\frac{1-x}{x}\right)^{3/2}\left[1 + x^{3/2}\, \zeta\left(\frac{3}{2},1+x\right)\right],
\label{t}
\ee
where
\bb
x = 1\!-\! P^{7/2} = 1-\left(\frac{b}{b_c}\right)^7 
= 1-\frac{384}{85\pi}\left(\frac{M}{4\mu}\right)
\left(\frac{v}{c}\right)^2\left(\frac{cvb}{2GM}\right)^7
\label{x}
\ee
is a measure for how much the impact parameter $b$ is less than the critical impact parameter $b_c = [(85\pi/384)(4\mu/M)]^{1/7}(2GM/c^2)(v/c)^{-9/7}$.

For a simple elementary function approximation for the dimensionless rescaled time $t(P)$ using 
\bb
z \equiv \zeta(3/2) \approx 2.612\,375,
\label{z}
\ee
one can use the approximation for $0 \leq x \leq 1$ that
\bb
\zeta\left(\frac{3}{2},1+x\right) \approx \zeta_a(x) \equiv z - \frac{2x}{1+x},
\label{za}
\ee
which in this range of $x = 1 - P^{7/2} = 1 - (b/b_c)^7$ gives a maximum value of $\zeta_a(x)/\zeta(3/2,1+x)$ of 1.000\,161 at $x = 0.075\,253$ and a minimum of 0.998\,607 at $x = 0.631\,457$, so the approximation $\zeta_a(x)$ is always within about 0.14\% of $\zeta(3/2,1+x)$.  Then this gives
\bb
t(P) \approx t_a(P) = \left(\frac{P^{7/2}}{1-P^{7/2}}\right)^{3/2}\left[1 + (1-P^{7/2})^{3/2}\left(z-\frac{1-P^{7/2}}{1-0.5P^{7/2}}\right)\right],
\label{ta}
\ee
which gives $t_a(P)/t(P)$ with a minimum value of 0.999\,310 at $P=0.714\,950$ and a maximum value of 1.000\,010\,922 at $P=0.966\,481$, so the elementary function $t_a(P)$ (though with the nonelementary constant $z = \zeta(3/2)$ calculated numerically) is always within 0.07\% of $t(P)$.

After doing this calculation of the inspiral merger time, I found many papers
\cite{Lee1993,OLeary:2008myb,Clesse:2016ajp,Kovetz:2016kpi,Ali-Haimoud:2017rtz,Gondan:2017wzd,Rodriguez:2018pss,Raidal:2018bbj,Sasaki:2018dmp,Vaskonen:2019jpv,Korol:2019jud,Jedamzik:2020ypm,Kritos:2020fjw,Kocsis2022}
that used Eq.\ (5.14) of \cite{Peters:1964zz},
or the unnumbered last equation of that paper by Peters that is the approximation for initial eccentricity $e_0$ near 1 (equivalent to my $f_1 \equiv 1-e_1^2 \ll 1$ for the first bound orbit)
that is an adiabatic approximation for the inspiral merger time, which is not valid, since the relative changes in $E$, $a$, and $\tau$ from one orbit to the next are not small.  I have given a fuller discussion of this is in \cite{Page2024b} but shall not do so here.

\section{Cumulative Conditional Probability $P$\\
as a Function of the Inspiral Time $T=T_0t$}

The inspiral time, given by Eq.\ (\ref{T}), has the form $T = T_0(M,v)t(P)$ with dimensional timescale factor $T_0(M,v) = 2\pi GM/v^3$ depending only on the total mass $M = M_1 + M_2$ and the incoming relative velocity $v$, and the dimensionless rescaled time $t(P) = P^{21/4}\zeta(3/2,1\!-\! P^{7/2})$ depending only on the conditional cumulative probability $P = (b/b_c)^2$ for impact parameters smaller than $b$, given the condition that the impact parameter is smaller than the critical impact parameter $b_c = [(85\pi/384)(4\mu/M)]^{1/7}(2GM/c^2)(v/c)^{-9/7}$ that depends not only on the total mass $M = M_1 + M_2$ and the incoming relative velocity $v$ but also on the reduced mass $\mu = M_1M_2/M$.  This dimensionless rescaled time $t(P)$ can be well approximated by $t_a(P)$ given by Eq.\ (\ref{ta}) with the constant $z = \zeta(3/2) \approx 2.612\,375$.

However, it is not quite so simple to get a good approximate inverse of $t(P)$, namely, one for $P(t)$.  Of course, since $P = (1-x)^{2/7}$ and
\bb
t(P) \approx t_a(P(x)) = \left(\frac{1-x}{x}\right)^{3/2}\left[1+x^{3/2}\left(z-\frac{2x}{1+x}\right)\right],
\label{tax}
\ee
one can first look for approximations for $x(t)$ to use in $P(t) = [1-x(t)]^{2/7}$.  One can readily convert Eq.\ (\ref{tax}) to give the iteration
\bb
x_{i+1}(t) = \left\{1+t^{2/3}\left[1+x_i^{3/2}\left(z-\frac{2x_i}{1+x_i}\right)\right]^{-2/3}\right\}^{-1},
\label{xi+1}
\ee
Starting with $x_0(t)=0$ on the right hand side, we get
\bb
x_1(t) = (1+t^{2/3})^{-1},
\label{x1}
\ee
\bb
x_2(t) = \left\{1+t^{2/3}\left[1+(1+t^{2/3})^{-3/2}\left(z-2/(2+t^{2/3})\right)\right]^{-2/3}\right\}^{-1}.
\label{x2}
\ee

Alternatively, one can convert Eq.\ (\ref{ta}) to the iteration
\bb
P_{i+1}(t) = \left\{1+t^{-2/3}\left[1+(1-P_i^{7/2})^{3/2}\left(z-(1-P_i^{7/2})/(1-P_i^{7/2}/2)\right)\right]^{2/3}\right\}^{-2/7},
\label{Pi+1}
\ee
which with $P_0(t)=1$ gives
\bb
P_1(t) = (1+t^{-2/3})^{-2/7},
\label{P1}
\ee
\bb
P_2(t) = \left\{1+t^{-2/3}\left[1+(1+t^{2/3})^{-3/2}\left(z-2/(2+t^{2/3})\right)\right]^{2/3}\right\}^{-2/7}.
\label{P2}
\ee

Using the exact Eq.\ (\ref{t}) for $t(P)$ (exact in the limit that the number of orbits before coalescence becomes arbitrarily large) then gives $P_2(t(P))-P$ always nonnegative, going to zero at both $P=0$ and $P=1$, and having a maximum value of 0.009\,653 at $P=0.717\,477$, so the absolute error of $P_2(t)$ is always less then 1\%, and the rms error over all $0\leq P \leq 1$ is 0.004\,910, less than 0.5\%.  Alternatively, $P_2(t(P))/P$ is always greater than or equal to 1, approaching 1 both as $P$ approaches 0 and as $P$ approaches 1, and having a maximum value of 1.013\,788 at $P = 0.681\,169$, so the maximum relative error of $P_2(t)$ is nearly 1.4\%.

On the other hand, $t(P_2(t))-t$ is also always nonnegative, and it approaches 0 for both $t\rightarrow 0$ and $t\rightarrow\infty$, but its maximum value is 0.198\,467 for $t=5.998\,613$, and $t(P_2(t))/t$ is greater than 1 for $0<P<1$ and approaches 1 for both $t\rightarrow 0$ and $t\rightarrow\infty$, and its maximum value is 1.093\,884 for $t=0.690\,563$, so $t(P_2(t))$ has considerably more absolute and relative error than $P_2(t(P))$.

One can get some improvement by taking one more iteration to get $x_3(t)$ and $P_3(t)$, though their expressions as single formulas are too long to be worth writing out here.  I found that $P_3(t(P))-P$ even more rapidly approaches zero at both $P=0$ and $P=1$ and is generally positive in between, with a maximum value of 0.001\,474 at $P=0.767\,978$, though it crosses zero at $P=0.992\,748$ and becomes very slightly negative for values of $P$ between this crossing and $P=1$.  The extremely-tiny-magnitude minimum value is $-0.000\,000\,000\,817\,963$ for $P=0.994\,276$.  The rms value for this error over all $0\leq P \leq 1$ is 0.000\,649, which is itself very small.  The maximum value for $P_3(t(P))/P$ is 1.001\,946 at $P=0.750\,739$, less than 0.2\% error, and the minimum value for this ratio is 0.999\,999\,999\,171\,327 at $P=0.994\,274$.

Again $t(P_3(t))-t$ has a rather larger error, ranging from a maximum of 0.032\,243 at $t \approx 4.00108$ to a tiny-magnitude minimum of $-0.000\,103$ at $t \approx 600$, and $t(P_3(t))/t$ has a maximum of 1.014\,456 at $t\approx 1.1968$ and a minimum of 0.999\,999\,777 at $t\approx 390$.  Therefore, if one wants the maximum error of $t(P_i(t))-t$ and of $t(P_i(t))/t-1$ to be less than 1\%, one would need to go to at least one more iteration, to $i \geq 4$.

\section{Probability Density $dP/d\ln{t}$ per Logarithmic\\
Interval $d\ln{T}=d\ln{t}$ of the Inspiral Time $T=T_0t$}

By taking the inspiral time $T$ and dividing by the timescale factor $T_0 = 2\pi GM/v^3$ to get the dimensionless time $t = T/T_0 = T/(2\pi GM/v^3) = P^{21/4}\zeta(3/2,1-P^{7/2})$ by Eq.\ (\ref{T}), we can then differentiate the logarithm of this with respect to the cumulative conditional probability $P=(b/b_c)^2$ that the impact parameter is less than $b$, given that is it less than the critical impact parameter $b_c$ for capture of the initially unbound black holes of total mass $M = M_1+M_2$, reduced mass $\mu = M_1M_2/M$, and relative velocity $v$ (all of which go into Eq.\ (\ref{bc}) for $b_c$) by the emission of gravitational radiation.  Then the reciprocal of this logarithmic derivative gives the probability density per logarithmic interval of inspiral time $T$, which for fixed $M$, $\mu$, and $v$ (but allowing $b$ and hence $P$, $t$, and $T=T_0t = (2\pi GM/v^3)t$ to vary) gives
\bb
\rho \equiv \frac{dP}{d\ln{T}} = t\frac{dP}{dt} = \frac{4}{21}P
\left[1 + \frac{P^{7/2}\zeta(5/2,1-P^{7/2})}{\zeta(3/2,1-P^{7/2})}\right]^{-1}.
\label{rho}
\ee

Next, let me give approximate formulas for $\rho$ as a function of $t$ instead of the exact (within my approximation of a large number of inspiral orbits) Eq.\ (\ref{rho}) as a function of the cumulative probability $P$.  Using $x = 1-P^{7/2}$ from Eq.\ (\ref{x}), Eq.\ (\ref{T}) gives $t = (1-x)^{3/2}\zeta(3/2,x)$, and then the approximation of Eq.\ (\ref{za}) with $z = \zeta(3/2)$ gives, for $0\leq x\leq 1$ as is the case here,
\bb
\zeta\left(\frac{3}{2},x\right) \approx x^{-3/2} + z - 2 + \frac{2}{1+x}.
\label{zea}
\ee
This leads to the further approximation
\bb
\zeta\left(\frac{5}{2},x\right) = -\frac{2}{3}\frac{d}{dx}\zeta\left(\frac{3}{2},x\right) \approx x^{-5/2}+\frac{4/3}{(1+x)^2}.
\label{zeb}
\ee

It is convenient to define the following factors which, for $0 \leq P \leq 1$ or $0 \leq x = 1-P^{7/2} \leq 1$, remain within a factor of $z = \zeta(3/2) \approx  2.612$ of unity:
\bb
F \equiv x^{3/2}\zeta(3/2,x) = 1 + x^{3/2}\zeta(3/2,1+x),
\label{F}
\ee
\ba
G &\equiv& \frac{4}{21}\frac{(1-P^{7/2})P}{\rho} 
  = x+\frac{x(1-x)\zeta(5/2,x)}{\zeta(3/2,x)} \nonumber \\
  &=& x + \frac{(1-x)[1+x^{5/2}\zeta(5/2,1+x)]}{1+x^{3/2}\zeta(3/2,1+x)}.
\label{G}
\ea

Using the approximations of Eqs.\ (\ref{zea}) and (\ref{zeb}), these give
\bb
F \approx F_a \equiv 1 + x^{3/2}\left(z - \frac{2x}{1+x}\right),
\label{Fa}
\ee
\bb
G \approx G_a \equiv x + \frac{(1-x)\left[1+x^{5/2}\frac{(4/3)}{(1+x)^2}\right]}
{1+x^{3/2}\left(z-\frac{2x}{1+x}\right)}.
\label{Ga}
\ee

It is also useful to define
\bb
y \equiv \frac{1-x}{x} \equiv \frac{P^{7/2}}{1-P^{7/2}} 
\equiv\left(\frac{t}{F}\right)^{2/3},
\label{y}
\ee
so that
\bb
P = (1-x)^{2/7} = \left(\frac{y}{1+y}\right)^{2/7}
  = \frac{t^{4/21}}{(F^{2/3}+t^{2/3})^{2/7}},
\label{Py}
\ee
\bb
x = 1-P^{7/2} = \frac{1}{1+y} = \frac{1}{1+(t/F)^{2/3}}.
\label{xy}
\ee
Replacing $F$ on the right hand side by $F_a(x)$ then leads to the iteration of Eq.\ (\ref{xi+1}), repeated here for convenience:
\bb
x_{i+1}(t) = \left\{1+t^{2/3}\left[1+x_i^{3/2}\left(z-\frac{2x_i}{1+x_i}\right)\right]^{-2/3}\right\}^{-1}.
\label{xj+1}
\ee
Then defining $F_{i+1} \equiv F_a(x_{i+1})$ leads to the following iteration for getting better and better approximations for $F(t)$:
\bb
F_{i+1}(t) = 1+\left[1+\left(\frac{t}{F_i}\right)^{2/3}\right]^{-3/2}
\left[z - \frac{2}{2+(t/F_i)^{2/3}}\right].
\label{Fi+1}
\ee
Starting with $F_0(t) = 1$, one gets
\bb
F_1(t) = 1+(1+t^{2/3})^{-3/2}\left(z-\frac{2}{2+t^{2/3}}\right),
\label{F1}
\ee
\ba
F_2(t)=1&\!\!\!\!+\!\!\!\!&\left[1+\left(\frac{t}{1+(1+t^{2/3})^{-3/2}\left(z-\frac{2}{2+t^{2/3}}\right)}\right)^{2/3}\right]^{-3/2}\times \nonumber \\
&& \left\{z-\frac{2}{2+t^{2/3}\left[1+(1+t^{2/3})^{-3/2}\left(z-\frac{2}{2+t^{2/3}}\right)\right]^{-2/3}}\right\}.
\label{F2}
\ea

One can similarly define a series of increasingly accurate approximation for $G(t)$ as $G_i(t) = G_a(x_i(t))$.  For example, $G_0(t) = 1$, and
\bb
G_1(t) = \frac{1}{1+t^{2/3}}\left[1+\frac{t^{2/3}(1+t^{2/3})^{3/2}
+\frac{4}{3}t^{2/3}\frac{(1+t^{2/3})}{(2+t^{2/3})^2}}
{(1+t^{2/3})^{3/2}+z-\frac{2}{2+t^{2/3}}}\right].
\label{G1}
\ee

From these approximations $F_i(t)$ and $G_i(t)$ for $F(t)$ and $G(t)$, one can get the following approximations for $\rho(t) \equiv dP/d\ln{T} = tdP/dt$: 
\bb
\rho_i(t) = \frac{4}{21}\left(\frac{t}{F_i(t)}\right)^{4/21}
\left[1+\left(\frac{t}{F_i(t)}\right)^{2/3}\right]^{-9/7}\frac{1}{G_i(t)}.
\label{Gi}
\ee
For example,
\bb
\rho_0(t) = \frac{4}{21}t^{4/21}(1+t^{2/3})^{-9/7},
\label{rho0}
\ee
\ba
\rho_1(t)&=&\frac{4}{21}t^{4/21}(1+t^{2/3})\left[1+(1+t^{2/3})^{-3/2}\left(z-\frac{2}{2+t^{2/3}}\right)\right]^{-4/21}\times \nonumber \\
&&\left\{1+t^{2/3}\left[1+(1+t^{2/3})^{-3/2}\left(z-\frac{2}{2+t^{2/3}}\right)\right]^{-2/3}\right\}^{-9/7}\times \nonumber \\
&&\left\{1+\frac{t^{2/3}\left[1+(1+t^{2/3})^{-5/2}\,\frac{4}{3}
\left(\frac{1+t^{2/3}}{2+t^{2/3}}\right)^2\right]}
{1+(1+t^{2/3})^{-3/2}\left(z-\frac{2}{1+t^{2/3}}\right)}\right\}^{-1}.
\label{rho1}
\ea
The explicit single formulas for the quantities with larger values of the subscript $i$ are too long to be worth writing out here, but they were evaluated by Mathematica for $i$ up through 3.

Mathematica was then used to examine the maximum relative errors of the various quantities, first for $F_a(P)$ and $G_a(P)$ as approximations for $F(P)$ and $G(P)$.

$F_a(P)/F(P)$ has the value 1 at both $P=0$ and at $P=1$.  This ratio has a maximum value of 1.000\,010\,922 at $P\approx 0.9666$, $x\approx 0.1121$, and $t\approx 24.3$, and it has a minimum value of 0.999\,309\,996 at $P\approx 0.71496$, $x\approx 0.6910$, and $t\approx 0.6078$.  Thus $F_a(P)$, given by Eq.\ (\ref{Fa}) in terms of $x=1-P^{7/2}$, always has a relative error less than 0.07\%, much less than the relative errors of $F_i(t)$ given below for $1\leq i \leq 3$.

$G_a(P)/G(P) = 1$ also at both $P=0$ and at $P=1$.  This ratio has a local minimum of 0.999\,993 at $P\approx 0.97777$, $x\approx 0.076$, and $t\approx 44.7$, a maximum of 1.000\,384 at $P\approx 0.85$, $x\approx 0.43$, and $t\approx 2.4$, and a global minimum of 0.999\,794 at $P\approx 0.570$, $x\approx 0.861$, and $t\approx 0.153$.  Thus $G_a(P)$, given by Eq.\ (\ref{Ga}) in terms of $x=1-P^{7/2}$, always has a relative error less than 0.04\%, again much less than the relative errors of $G_i(t)$ given below for $1\leq i \leq 3$.

Next, let us examine the maximum relative errors of approximations for quantities as explicit functions of the dimensionless inspiral time $t = T/T_0 = T/(2\pi GM/v^3)$, 
say $Q_i(t)$, relative to the exact expressions $Q(P)$ by using Mathematica to evaluate $Q_i(t(P))/Q(P)$ with the exact $t(P) = P^{21/4}\zeta(3/2,1-P^{7/2})$.

$F_1(t(P))/F(P) = 1$ at $P=0$ and at $P=1$, with a minimum value of 0.900\,704 at $P\approx 0.746$, $x\approx 0.642$, and $t\approx 0.81$, so $F_1(t)$ has a relative error up to nearly 10\%.

$F_2(t(P))/F(P) = 1$ at $P=0$ and at $P=1$, with a minimum value of 0.982\,994 at $P\approx 0.793$, $x\approx 0.556$, and $t\approx 1.27$, so $F_2(t)$ has a relative error up to just over 1.3\%.

$F_3(t(P))/F(P) = 1$ at $P=0$ and at $P=1$, with a minimum value of 0.996\,578 at $P\approx 0.804$, $x\approx 0.534$, and $t\approx 1.42$, so $F_3(t)$ has a relative error always less than 0.4\%.

$\rho_1(t(P))/\rho(P)$ has its maximum value of $z^{4/21} = 1.200\,702$ at $P=0$, $x=1$, and $t=0$, and this ratio goes to 1 at $P=1$, $x=0$, and $t=\infty$.  It has its minimum value 0.856\,112 at $P\approx 0.869$, $x\approx 0.388$, and $t\approx 2.97$.  Thus its relative error ranges from $+20\%$ down to nearly $-15\%$, which is not very good.

$\rho_2(t(P))/\rho(P)$ does go to 1 at both endpoints ($P=0$ and $P=1$).  This ratio has its maximum value 1.007\,634 at $P\approx 0.553$, $x\approx 0.874$, and $t\approx 0.130$, and it has its minimum value 0.975\,180 at $P\approx 0.868$, $x\approx 0.391$, and $t\approx 2.92$.  Thus its relative error is always a smaller than 2.5\%.

$\rho_3(t(P))/\rho(P)$ also goes to 1 at both endpoints ($P=0$ and $P=1$).  This ratio has its minimum value 0.995\,234 at $P\approx 0.864$, $x\approx 0.401$, and $t\approx 2.78$, and it has its maximum value 1.000\,001\,577 at $P\approx 0.9911$, $x\approx 0.031$, and $t\approx 180$.  Thus its relative error is always smaller than 0.5\%.

In summary, when I used the 3rd iteration, $i=3$ for $P_3(t)$ and $\rho_3(t)$, the relative error is always significantly smaller than 1\%, so for practical purposes these approximations should be good enough.  The explicit single formulas for $P_3(t)$ and $\rho_3(t)$ are too long to be copied here, but by the iterative formulas given above, they can readily be calculated and evaluated by Mathematica.

\section{Data for $t$, $T$, and $\rho = tdP/dt$ as Functions of $P$, and a Graph for $P$ and $\rho$ as Functions of $\log{t}$}

\begin{tabular}{||l|l|l|l||} \hline
$P=\left(\frac{b}{b_c}\right)^2$ & $t\!=\!P^{\frac{21}{4}}\,\zeta\left(\frac{3}{2},1\!-\! P^{\frac{7}{2}}\right)$ & $T$ for $10\,M_\odot$, 100 km/s 
& $\!\rho\!=\!t\frac{dP}{dt}\!\!=\!\!
\frac{4P}{21}\!\left(\!\!1\!+\!\frac{P^{\frac{7}{2}}\zeta(\frac{5}{2},1-P^{\frac{7}{2}})}{\zeta(\frac{3}{2},1-P^{\frac{7}{2}})}\!\!\right)^{\!\!\!-1\!}\!\!$ \\ \hline
0 & 0 & $\sim 2\pi GM/c^3 \sim 10^{-11}$ yr & 0 \\ \hline
0.01 &	$8.261\,057\times 10^{-11}$ & direct coalescence 
&0.001\,904\,762 \\ \hline
0.02 &	$3.143\,717\times 10^{-9}$ & direct coalescence 
& 0.003\,809\,522\\ \hline
0.05 & $3.860\,446\times 10^{-7}$ & nearly direct coalescence
& 0.009\,523\,673 \\ \hline
0.10 &	$1.469\,405\times 10^{-5}$ & $4\times 10^{-6}$ yr, $\beta_p^2 \sim 1/10$
& 0.019\,044\,525  \\ \hline
0.15 & $1.235\,812\times 10^{-4}$ & $3\times 10^{-5}$ yr, $\beta_p^2 \sim 1/15$
& 0.028\,552\,230  \\ \hline
0.20 & $5.605\,866\times 10^{-4}$ & $1.5\times 10^{-4}$ yr, $\beta_p^2 \sim 1/20$
& 0.038\,025\,045  \\ \hline
0.25 & $1.814\,879\times 10^{-3}$ &	$4.8\times 10^{-4}$ yr, $\beta_p^2 \sim 1/25$
& 0.047\,426\,785 \\ \hline
0.30 & $4.752\,460\times 10^{-3}$ &	$1.3\times 10^{-3}$ yr, $\beta_p^2 \sim 1/30$
& 0.056\,703\,625 \\ \hline
0.35 & $1.076\,525\times 10^{-2}$ & $2.8\times 10^{-3}$ yr, $\beta_p^2 \sim 1/35$
& 0.065\,780\,052 \\ \hline
0.40 & $2.196\,685\times 10^{-2}$ & $5.8\times 10^{-3}$ yr, $\beta_p^2 \sim 1/40$
& 0.074\,553\,181 \\ \hline
0.45 & $4.146\,907\times 10^{-2}$ & $1.10\times 10^{-2}$ yr, $\beta_p^2 \sim 1/45$
& 0.082\,884\,298 \\ \hline
0.50 & $7.380\,184\times 10^{-2}$ & $1.95\times 10^{-2}$ yr, $\beta_p^2 \sim 1/50$
& 0.090\,586\,038 \\ \hline
0.55 & $1.255\,964\times 10^{-1}$ & $3.32\times 10^{-2}$ yr, $\beta_p^2 \sim 1/55$
& 0.097\,402\,690 \\ \hline
0.60 & $2.067\,776\times 10^{-1}$ & $5.46\times 10^{-2}$ yr, $\beta_p^2 \sim 1/60$
& 0.102\,979\,918 \\ \hline
0.65 & $3.328\,098\times 10^{-1}$ &	$8.79\times 10^{-2}$ yr, $\beta_p^2 \sim 1/65$
& 0.106\,818\,463 \\ \hline
0.70 & $5.293\,246\times 10^{-1}$ & $1.40\times 10^{-1}$ yr, $\beta_p^2 \sim 1/70$
& 0.108\,205\,199 \\ \hline
0.75 & $8.427\,883\times 10^{-1}$ & $2.23\times 10^{-1}$ yr, $\beta_p^2 \sim 1/75$
& 0.106\,118\,525 \\ \hline
0.80 &	1.369\,021 & $3.62\times 10^{-1}$ yr, $\beta_p^2 \sim 1/80$
& 0.099\,129\,628 \\ \hline
0.85 & 2.347\,172 & $6.20\times 10^{-1}$ yr, $\beta_p^2 \sim 1/85$
& 0.085\,418\,810 \\ \hline
0.90 & 4.590\,029 &	1.21 years, $\beta_p^2 \sim 1/90$
& 0.063\,315\,302 \\ \hline
0.95 & $1.324\,718\times 10^1$ & 3.50 years, $\beta_p^2 \sim 1/95$
& 0.033\,208\,069 \\ \hline
0.98 & $5.265\,574\times 10^1$ & $1.39\times 10^1$ years, $\beta_p^2 \sim 0.01$
& 0.013\,250\,261 \\ \hline
0.99 & $1.500\,313\times 10^2$ & $3.96\times 10^1$ years, $\beta_p^2 \sim 0.01$
& 0.006\,619\,004 \\ \hline
0.999 &	$4.815\,761\times 10^3$ & $1.27\times 10^3$ years, $\beta_p^2 \sim 0.01$
& $6.655\,242\times 10^{-4}$ \\\hline
0.999\,9 & $1.526\,718\times 10^5$ & $4.03\times 10^4$ years, $\beta_p^2 \sim 0.01$
& $6.665\,281\times 10^{-5}$ \\ \hline
0.999\,99 &	$4.829\,293\times 10^6$ & $1.28\times 10^6$ years, $\beta_p^2 \sim 0.01$
& $6.666\,520\times 10^{-6}$ \\ \hline
0.999\,999 & $1.527\,202\times 10^8$ & $4.04\times 10^7$ years, $\beta_p^2 \sim 0.01$
& $6.666\,652\times 10^{-7}$ \\ \hline
0.999\,999\,9 & $4.829\,451\times 10^9$ & $1.28\times 10^9$ years, $\beta_p^2 \sim 0.01$
& $6.666\,665\times 10^{-8}$ \\ \hline
0.999\,999\,99 & $1.527\,207\times 10^{11}$ & $4.04\times 10^{10}$ years, $\beta_p^2 \sim 0.01$
& $6.666\,666\times 10^{-9}$ \\ \hline
0.999\,999\,999 &$ 4.829\,452\times 10^{12}$ & $1.28\times 10^{12}$ years, $\beta_p^2 \sim 0.01$
& $6.666\,667\times 10^{-10}$ \\ \hline
\end{tabular}

\vspace{2mm}

Here the first column gives the cumulative conditional probability $P = (b/b_c)^2$ where $b_c$, given by Eq.\ (\ref{bc}), is the critical impact parameter for capture, the second column gives the dimensionless inspiral merger time $t = T/T_0 = T/(2\pi GM/v^3) \approx P^{21/4}\zeta(3/2,1-P^{7/2})$, the third column giving the inspiral time $T$ for $M_1 = M_2 = 5 M_\odot$ (so that $M = 10 M_\odot$ and $\eta = 1/4$) and for $v = 100$\ km/s $\approx (1/3)10^{-3}c$ (as well as estimates for $\beta_p^2 = (v_p/c)^2 \sim 2GM/r_p$ at periapsis), and the fourth column gives the probability density per logarithmic interval of the inspiral time $T$, namely $\rho = dP/d\ln{T} = dP/d\ln{t}$.  Note that with $M_1=M_2$ and this value of $v$, one gets at periapsis $\beta_p^2 \equiv v_p/c \sim 0.01/P$, which is not very small when $P$ is small, giving uncalculated relative errors for $T$ of the order of $\beta_p^2$.  Thus the full 7 digits given for $t = T/T_0$ in the second column and for $\rho = dP/d\ln{T}$ in the fourth column are only all relevant for $\beta_p^2 \sim [\beta^2/(4\mu/M)]^{2/7}/P \stackrel{<}{\sim} 10^{-6}$ or $\beta \sim (4\mu/M)^{1/2}P^{7/4}\beta_p^{7/2} \stackrel{<}{\sim} 10^{-11}(4\mu/M)^{1/2} P^{7/4}$, but they are given for completeness.

\begin{figure}
\includegraphics{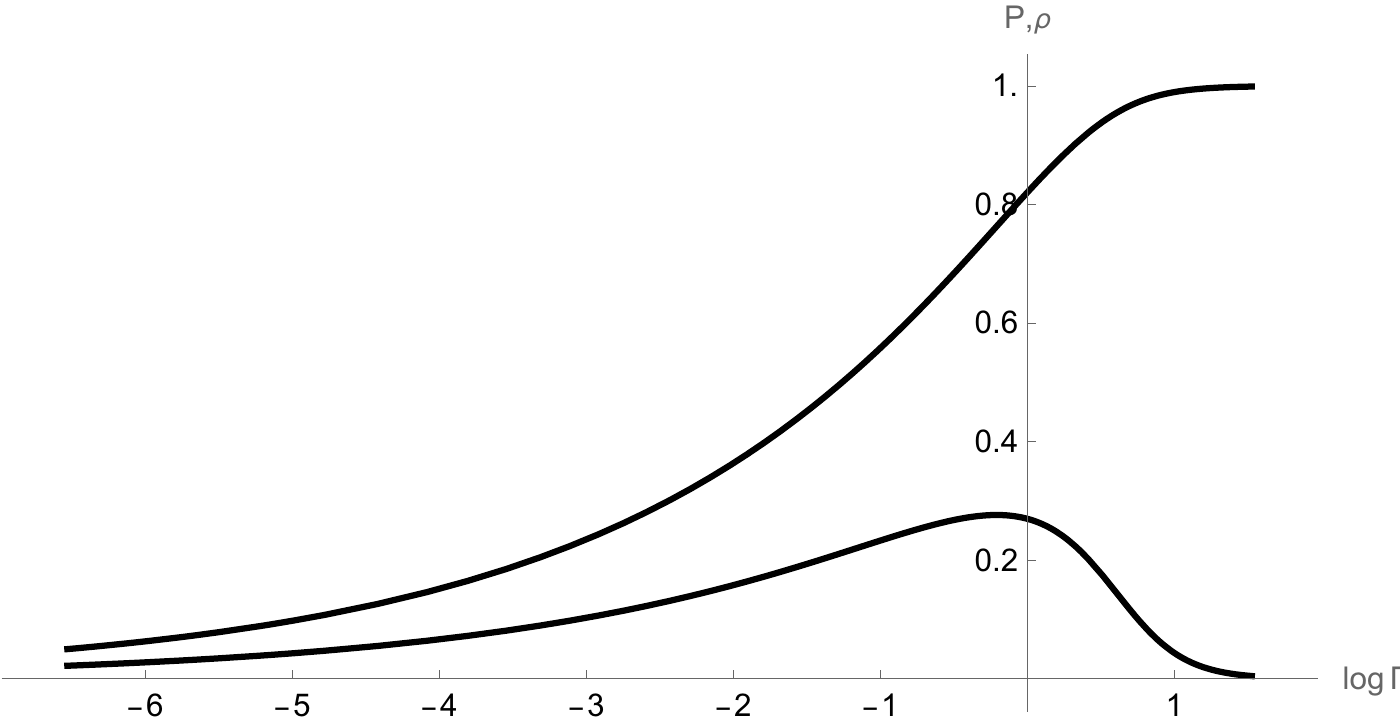}
\caption{The top graph gives the cumulative conditional probability $P = (b/b_c)^2$ for impact parameters smaller than $b$ given that they are smaller than the critical impact parameter $b_c = [340\pi G^7 M^6 \mu/(3 c^5 v^9)]^{1/7}$ for capture into bound orbits, and the bottom graph gives $(\ln{10})*\rho = dP/d\log{t}$, both as functions of $\log{t}$, the common logarithm of the dimensionless inspiral time $t=T/(2\pi GM/v^3)$ with $T$ the inspiral time for two black holes of total mass $M$, reduced mass $\mu$, and incident relative velocity $v$.}
\end{figure}

As a function of the rescaled time 
$t=T/(2\pi GM/v^3) = P^{21/4}\zeta(3/2,1-P^{7/2})$,
the probability distribution tabulated in the Table and shown in Figure 1 is very broad, with $t$ varying by a factor of over a trillion ($1.816\times 10^{12}$) between the 1st and 99th percentiles (between $P=0.01$ and $P=0.99$), by a factor over 16 billion ($1.675\times 10^{10}$) between the 2nd and 98th percentile, by a factor over 34 million ($3.432\times 10^7$) between the 5th and the 95 percentile, by a factor over 312 thousand ($3.127\times 10^5$) between the 10th and 90th percentile, and by a factor slightly over 464 between the 25th and 75 percentiles (the first and third quantiles, $P=0.25$ and $P=0.75$).

The tails of the probability distribution per range of $\ln{t}$, $\rho = t dP/dt$, also decay slowly, asymptotically having the following form for very small and very large $t$ respectively, with $z \equiv \zeta(3/2) \approx 2.612\,375$:

\bb
\rho(t) = \frac{4}{21}\left(\frac{t}{z}\right)^{4/21}
\left[1-\frac{9}{7}\left(\frac{t}{z}\right)^{2/3}+O(t)\right],
\label{rhosmallt}
\ee
\bb
\rho(t) = \frac{4}{21}t^{-2/3}
\left[1-\frac{9}{7}t^{-2/3}+O(t^{-4/3})\right].
\label{rholarget}
\ee
These tails imply that the mean value of $t^s$ converges only for $-4/21 < s < 2/3$, though the first (left-hand) inequality in (\ref{binequalities}), impact parameter $b \gg 2GM/(cv)$, implies that the divergence at infinitesimal $t$ when $s < -4/21$ is not physical.  If one defines the inspiral time $T = (2\pi GM/v^3)$ so that it is always greater than $2GM/c^3$, the light travel time for a distance equal to the Schwarzschild radius corresponding to the total mass $M$, this would imply that one always has $\pi t > (v/c)^3$, which can be very small, but not infinitesimal, thus cutting off the divergence arising from the approximation $t = P^{21/4}\zeta(3/2,1-P^{7/2})$ that only applies for $t \gg (v/c)^3$.  

On the other hand, the divergence at large $t$ for $s > 2/3$ would be real under the idealization made here that the black hole pair is isolated in otherwise flat spacetime.  However, in reality there would be other matter around to perturb the black hole orbits and presumably put a highly situation-dependent upper limit on the inspiral time $T$, which when $t \gg 1$ would be dominated by the initial orbit of eccentricity very near one and hence a huge semimajor axis that is much larger than the distance of closest approach near which just barely enough gravitational radiation occurs to produce a very weakly bound orbit.  It shall be left to future research to estimate the upper limits on $t$ from the gravitational perturbations of other objects in realistic environments of the pair of black holes.

However, if for simplicity we continue to focus on the idealized case of two isolated black holes approaching each other with relative velocity $v << c$ with an impact parameter $b$ less than the critical impact parameter $b_c$ for capture into a bound orbit, and if we assume a uniform distribution of impact parameter vectors in the plane perpendicular to the incident relative velocity (so that the probability that the magnitude of that vector in the plane is less than $b$, given that it is less than $b_c$, is $P = (b/b_c)^2$ with the exponent 2 coming from the dimension of the plane, then for $b \gg 2GM/(cv)$ we do get $t = P^{21/4}\zeta(3/2,1-P^{7/2})$, which leads to a divergence in the mean value of $t$ (as well as for the mean value of any positive power of $t$ greater than or equal to 2/3).

Taking the simplest fractional power less than 2/3, I found that the mean value of $t^{1/2}$ is approximately 1.253, whose square is 2.570, though the standard deviation of the probability distribution for $t^{1/2}$ diverges.

A quantity both of whose mean and standard deviation is finite is the logarithm of $t$, which has a mean value of $-3.899\,276$.  (This is for the natural logarithm; the common logarithm, with base 10, used in Figure 1, has a mean value of $-1.693\,630$.)  The mean value of $\ln^2{t}$ is 46.977\,643, leading to a standard deviation of $\ln{t}$ of 5.636\,468.  The exponential of the mean of $\ln{t}$ is 0.020\,0247, which is small, even smaller than the median value of $t$ of 0.073\,802 (at $P = 0.5$), but the exponential of the standard deviation of $\ln{t}$ is 280.470, another consequence of the breadth of the probability distribution for $t$ and $\ln{t}$.

Besides the rather long approximate formulas given in the previous section for $P(t)$ and $\rho(t) = t dP/dt$, it may be useful to have some shorter approximate formulas, which I shall present here.  I shall start with 
\bb
F = (P^{-7/2}-1)^{3/2} t = (1-P^{7/2})^{3/2}\zeta(3/2,1-P^{7/2}),
\label{Pdef}
\ee
which varies in value from $z \equiv \zeta(3/2)$ at $P = 0$ (which gives $t=0$) to a value of 1 at $P = 1$ (with gives $t = \infty$).  For large $t$, $F(t)$ has the approximate asymptotic form $F(t) = 1+z/t+O(1/t^2)$, so a simple fit is given by the rational function that I shall call $F_b(t)$:
\bb
F_b(t) = \frac{z^2 + (z-1)t}{z + (z-1)t}.
\label{Fb}
\ee

This then leads to the following approximate expression for $P(t)$:
\bb
P_b(t) = \left[1+\left(\frac{F_b}{t}\right)^{2/3}\right]^{-2/7}
       = \left[1+\left(\frac{z^2 + (z-1)t}{zt + (z-1)t^2}\right)^{2/3}\right]^{-2/7}.
\label{Pb}
\ee
Next I shall give $P_b(t)$ in respective forms in which one can readily see the behavior for small or large $t$:
\bb
P_b(t) = \left(\frac{t}{z}\right)^{4/21}
\left(\frac{z^2+z(z-1)t}{z^2+(z-1)t}\right)^{4/21}
\left[1+\left(\frac{z+(z-1)t}{z^2+(z-1)t}\,t\right)^{2/3}\right]^{-2/7},
\label{Pbsmallt}
\ee
\bb
P_b(t) = \left[1+\left(\frac{z-1+z^2\, t^{-1}}{z-1+z\,t^{-1}}\,t^{-1}\right)^{2/3}\right]^{-2/7}.
\label{Pblarget}
\ee

Finally, one can explicitly differentiate these expressions to get $\rho_b(t) = t dP_b/dt$, again given respectively in two forms, the first more convenient for small $t$, and the second more convenient for large $t$:
\bb
\rho_b(t) = \frac{\frac{4}{21}\left(\frac{t}{z}\right)^{4/21}
\left(\frac{z^2+z(z-1)t}{z^2+(z-1)t}\right)^{25/21}
\left(1-\frac{(z-1)^3 t^2}{z[z+(z-1)t]^2}\right)}
{\left[1+\left(\frac{z+(z-1)t}{z^2+(z-1)t}t\right)^{2/3}\right]^{9/7}},
\label{rhobsmallt}
\ee
\bb
\rho_b(t) = \frac{\frac{4}{21}t^{-2/3}
\left(\frac{z-1+z\,t^{-1}}{z-1+z^2\, t^{-1}}\right)^{1/3}
\left(1+z(z-1)\frac{2(z-1)\,t^{-1} + z\,t^{-2}}{(z-1+z\,t^{-1})^2}\right)}
{\left[1+\left(\frac{z-1+z^2\,t^{-1}}{z-1+z\,t^{-1}}\,t^{-1}\right)^{2/3}\right]^{9/7}}.
\label{rhoblarget}
\ee

To examine the accuracy of the approximations $P_b(t)$ and $\rho_b(t)$, I used Mathematica to evaluate the ratios $P_b(t(P))/P$ and $\rho_b(t(P))/\rho(P)$ using the exact Eq.\ (\ref{t}) for $t(P)$ and Eq.\ (\ref{rho}) for $\rho(P)$.  The results showed that both ratios approached 1 at both $P=0$ and $P=1$, with the deviations decreasing faster than linearly.  $P_b(t(P))/P$ was always less than or equal to 1, with a minimum value of 0.990\,641 at $P \approx 0.653$ or $t\approx 0.34$, so its relative error was always less than 1\%.  Being a derivative of the approximate $P_b(t)$, $\rho_b(t) = t dP_b(t)$ had somewhat larger relative error, but still always less than 3.4\%.  In particular, $\rho_b(t(P)/\rho(P)$ had a minimum value of 0.978\,989 at $P\approx 0.524$ or $t\approx 0.0959$ and a maximum value of 1.033\,648 at $P\approx 0.895$ or $t\approx 4.26$.  

\section{Conclusions}

For two black holes of masses $M_1$ and $M_2$ (and hence total mass $M=M_1+M_2$ and reduced mass $\mu = M_1M_2/M$) in otherwise empty spacetime approaching each other at a nonrelativistic velocity $v \ll c$, as they pass each other, they will emit enough energy in gravitational radiation to become captured if their impact parameter $b$ is less than the critical impact parameter
$b_c = [340\pi G^7 M^6 \mu/(3 c^5 v^9)]^{1/7} 
= [(85\pi/384)(4\mu/M)]^{1/7}(2GM/c^2)(v/c)^{-9/7}$.
Assuming a uniform flux of black hole pairs approaching each other, the cumulative probability for impact parameters up to $b$, conditional upon them being less than $b_c$, is $P = (b/b_c)^2$.  If $2GM/(cv) \ll b < b_c$, the point of closest approach on the first pass is much larger than the Schwarzschild radius of each of the black holes, and the motion remains nonrelativistic, with many orbits at eccentricity near one and gradually decreasing semimajor axis and period.  Under these approximations, one gets a simple formula for the evolution of the orbits and the total time of inspiral as $T = (2\pi GM/v^3)P^{21/4}\zeta(3/2,1-P^{7/2})$ in terms of the Hurwitz zeta function.  From this, one can get a formula for $\rho = t dP/dt$, the probability per logarithmic interval of the inspiral time, as an explicit function of $P = (b/b_c)^2$.  Further formulas were given as approximate explicit closed-form algebraic functions for $t(P)$ and $\rho(P)$, as well as for the inverse function $P(t)$ and for $\rho(t)$.

\section*{Acknowledgments}

This work was motivated by research by Ashaduzzaman Joy on the number of gravitons emitted by black-hole coalescences observed by LIGO and Virgo.  Financial support was provided by the Natural Sciences and Engineering Research Council of Canada.


\vspace{3cm}

\begin{thebibliography}{}

\bibitem{LIGOScientific:2016aoc}
B.~P.~Abbott \textit{et al.} [LIGO Scientific and Virgo],
``Observation of Gravitational Waves from a Binary Black Hole Merger,''
Phys. Rev. Lett. \textbf{116}, no.6, 061102 (2016)
doi:10.1103/PhysRevLett.116.061102
[arXiv:1602.03837 [gr-qc]].

\bibitem{LIGOScientific:2016emj}
B.~P.~Abbott \textit{et al.} [LIGO Scientific and Virgo],
``GW150914: The Advanced LIGO Detectors in the Era of First Discoveries,''
Phys. Rev. Lett. \textbf{116}, no.13, 131103 (2016)
doi:10.1103/PhysRevLett.116.131103
[arXiv:1602.03838 [gr-qc]].

\bibitem{LIGOScientific:2016vbw}
B.~P.~Abbott \textit{et al.} [LIGO Scientific and Virgo],
``GW150914: First results from the search for binary black hole coalescence with Advanced LIGO,''
Phys. Rev. D \textbf{93}, no.12, 122003 (2016)
doi:10.1103/PhysRevD.93.122003
[arXiv:1602.03839 [gr-qc]].

\bibitem{LIGOScientific:2016vlm}
B.~P.~Abbott \textit{et al.} [LIGO Scientific and Virgo],
``Properties of the Binary Black Hole Merger GW150914,''
Phys. Rev. Lett. \textbf{116}, no.24, 241102 (2016)
doi:10.1103/PhysRevLett.116.241102
[arXiv:1602.03840 [gr-qc]].

\bibitem{LIGOScientific:2016lio}
B.~P.~Abbott \textit{et al.} [LIGO Scientific and Virgo],
``Tests of general relativity with GW150914,''
Phys. Rev. Lett. \textbf{116}, no.22, 221101 (2016)
[erratum: Phys. Rev. Lett. \textbf{121}, no.12, 129902 (2018)]
doi:10.1103/PhysRevLett.116.221101
[arXiv:1602.03841 [gr-qc]].

\bibitem{LIGOScientific:2016kwr}
B.~P.~Abbott \textit{et al.} [LIGO Scientific and Virgo],
``The Rate of Binary Black Hole Mergers Inferred from Advanced LIGO Observations Surrounding GW150914,''
Astrophys. J. Lett. \textbf{833}, no.1, L1 (2016)
doi:10.3847/2041-8205/833/1/L1
[arXiv:1602.03842 [astro-ph.HE]].

\bibitem{LIGOScientific:2016vpg}
B.~P.~Abbott \textit{et al.} [LIGO Scientific and Virgo],
``Astrophysical Implications of the Binary Black-Hole Merger GW150914,''
Astrophys. J. Lett. \textbf{818}, no.2, L22 (2016)
doi:10.3847/2041-8205/818/2/L22
[arXiv:1602.03846 [astro-ph.HE]].

\bibitem{LIGOScientific:2016sjg}
B.~P.~Abbott \textit{et al.} [LIGO Scientific and Virgo],
``GW151226: Observation of Gravitational Waves from a 22-Solar-Mass Binary Black Hole Coalescence,''
Phys. Rev. Lett. \textbf{116}, no.24, 241103 (2016)
doi:10.1103/PhysRevLett.116.241103
[arXiv:1606.04855 [gr-qc]].

\bibitem{LIGOScientific:2016dsl}
B.~P.~Abbott \textit{et al.} [LIGO Scientific and Virgo],
``Binary Black Hole Mergers in the first Advanced LIGO Observing Run,''
Phys. Rev. X \textbf{6}, no.4, 041015 (2016)
[erratum: Phys. Rev. X \textbf{8}, no.3, 039903 (2018)]
doi:10.1103/PhysRevX.6.041015
[arXiv:1606.04856 [gr-qc]].

\bibitem{LIGOScientific:2017bnn}
B.~P.~Abbott \textit{et al.} [LIGO Scientific and VIRGO],
``GW170104: Observation of a 50-Solar-Mass Binary Black Hole Coalescence at Redshift 0.2,''
Phys. Rev. Lett. \textbf{118}, no.22, 221101 (2017)
[erratum: Phys. Rev. Lett. \textbf{121}, no.12, 129901 (2018)]
doi:10.1103/PhysRevLett.118.221101
[arXiv:1706.01812 [gr-qc]].

\bibitem{LIGOScientific:2017ycc}
B.~P.~Abbott \textit{et al.} [LIGO Scientific and Virgo],
``GW170814: A Three-Detector Observation of Gravitational Waves from a Binary Black Hole Coalescence,''
Phys. Rev. Lett. \textbf{119}, no.14, 141101 (2017)
doi:10.1103/PhysRevLett.119.141101
[arXiv:1709.09660 [gr-qc]].

\bibitem{LIGOScientific:2017vox}
B.~P.~Abbott \textit{et al.} [LIGO Scientific and Virgo],
``GW170608: Observation of a 19-solar-mass Binary Black Hole Coalescence,''
Astrophys. J. Lett. \textbf{851}, L35 (2017)
doi:10.3847/2041-8213/aa9f0c
[arXiv:1711.05578 [astro-ph.HE]].

\bibitem{LIGOScientific:2018dkp}
B.~P.~Abbott \textit{et al.} [LIGO Scientific and Virgo],
``Tests of General Relativity with GW170817,''
Phys. Rev. Lett. \textbf{123}, no.1, 011102 (2019)
doi:10.1103/PhysRevLett.123.011102
[arXiv:1811.00364 [gr-qc]].

\bibitem{LIGOScientific:2018mvr}
B.~P.~Abbott \textit{et al.} [LIGO Scientific and Virgo].
``GWTC-1: A Gravitational-Wave Transient Catalog of Compact Binary Mergers Observed by LIGO and Virgo during the First and Second Observing Runs,''
Phys. Rev. X \textbf{9}, no.3, 031040 (2019)
doi:10.1103/PhysRevX.9.031040
[arXiv:1811.12907 [astro-ph.HE]].

\bibitem{LIGOScientific:2018jsj}
B.~P.~Abbott \textit{et al.} [LIGO Scientific and Virgo],
``Binary Black Hole Population Properties Inferred from the First and Second Observing Runs of Advanced LIGO and Advanced Virgo,''
Astrophys. J. Lett. \textbf{882}, no.2, L24 (2019)
doi:10.3847/2041-8213/ab3800
[arXiv:1811.12940 [astro-ph.HE]].

\bibitem{LIGOScientific:2019lzm}
R.~Abbott \textit{et al.} [LIGO Scientific and Virgo],
``Open data from the first and second observing runs of Advanced LIGO and Advanced Virgo,''
SoftwareX \textbf{13}, 100658 (2021)
doi:10.1016/j.softx.2021.100658
[arXiv:1912.11716 [gr-qc]].

\bibitem{LIGOScientific:2019fpa}
B.~P.~Abbott \textit{et al.} [LIGO Scientific and Virgo],
``Tests of General Relativity with the Binary Black Hole Signals from the LIGO-Virgo Catalog GWTC-1,''
Phys. Rev. D \textbf{100}, no.10, 104036 (2019)
doi:10.1103/PhysRevD.100.104036
[arXiv:1903.04467 [gr-qc]].

\bibitem{LIGOScientific:2020aai}
B.~P.~Abbott \textit{et al.} [LIGO Scientific and Virgo],
``GW190425: Observation of a Compact Binary Coalescence with Total Mass $\sim 3.4 M_{\odot}$,''
Astrophys. J. Lett. \textbf{892}, no.1, L3 (2020)
doi:10.3847/2041-8213/ab75f5
[arXiv:2001.01761 [astro-ph.HE]].

\bibitem{LIGOScientific:2020stg}
R.~Abbott \textit{et al.} [LIGO Scientific and Virgo],
``GW190412: Observation of a Binary-Black-Hole Coalescence with Asymmetric Masses,''
Phys. Rev. D \textbf{102}, no.4, 043015 (2020)
doi:10.1103/PhysRevD.102.043015
[arXiv:2004.08342 [astro-ph.HE]].

\bibitem{LIGOScientific:2020zkf}
R.~Abbott \textit{et al.} [LIGO Scientific and Virgo],
``GW190814: Gravitational Waves from the Coalescence of a 23 Solar Mass Black Hole with a 2.6 Solar Mass Compact Object,''
Astrophys. J. Lett. \textbf{896}, no.2, L44 (2020)
doi:10.3847/2041-8213/ab960f
[arXiv:2006.12611 [astro-ph.HE]].

\bibitem{LIGOScientific:2020iuh}
R.~Abbott \textit{et al.} [LIGO Scientific and Virgo],
``GW190521: A Binary Black Hole Merger with a Total Mass of $150  M_{\odot}$,''
Phys. Rev. Lett. \textbf{125}, no.10, 101102 (2020)
doi:10.1103/PhysRevLett.125.101102
[arXiv:2009.01075 [gr-qc]].

\bibitem{LIGOScientific:2020ufj}
R.~Abbott \textit{et al.} [LIGO Scientific and Virgo],
``Properties and Astrophysical Implications of the 150 M$_\odot$ Binary Black Hole Merger GW190521,''
Astrophys. J. Lett. \textbf{900}, no.1, L13 (2020)
doi:10.3847/2041-8213/aba493
[arXiv:2009.01190 [astro-ph.HE]].

\bibitem{LIGOScientific:2020ibl}
R.~Abbott \textit{et al.} [LIGO Scientific and Virgo],
``GWTC-2: Compact Binary Coalescences Observed by LIGO and Virgo During the First Half of the Third Observing Run,''
Phys. Rev. X \textbf{11}, 021053 (2021)
doi:10.1103/PhysRevX.11.021053
[arXiv:2010.14527 [gr-qc]].

\bibitem{LIGOScientific:2020tif}
R.~Abbott \textit{et al.} [LIGO Scientific and Virgo],
``Tests of general relativity with binary black holes from the second LIGO-Virgo gravitational-wave transient catalog,''
Phys. Rev. D \textbf{103}, no.12, 122002 (2021)
doi:10.1103/PhysRevD.103.122002
[arXiv:2010.14529 [gr-qc]].

\bibitem{LIGOScientific:2021usb}
R.~Abbott \textit{et al.} [LIGO Scientific and VIRGO],
``GWTC-2.1: Deep Extended Catalog of Compact Binary Coalescences Observed by LIGO and Virgo During the First Half of the Third Observing Run,''
[arXiv:2108.01045 [gr-qc]].

\bibitem{LIGOScientific:2021djp}
R.~Abbott \textit{et al.} [LIGO Scientific, VIRGO and KAGRA],
``GWTC-3: Compact Binary Coalescences Observed by LIGO and Virgo During the Second Part of the Third Observing Run,''
[arXiv:2111.03606 [gr-qc]].

\bibitem{LIGOScientific:2021sio}
R.~Abbott \textit{et al.} [LIGO Scientific, VIRGO and KAGRA],
``Tests of General Relativity with GWTC-3,''
[arXiv:2112.06861 [gr-qc]].

\bibitem{Ghosh:2022xhn}
A.~Ghosh [LIGO Scientific--Virgo--Kagra],
``Summary of Tests of General Relativity with GWTC-3,''
[arXiv:2204.00662 [gr-qc]].

\bibitem{Peters:1963ux}
P.~C.~Peters and J.~Mathews,
``Gravitational Radiation from Point Masses in a Keplerian Orbit,''
Phys. Rev. \textbf{131}, 435-439 (1963)
doi:10.1103/PhysRev.131.435.

\bibitem{Peters:1964qza}
P.~C.~Peters,
``Gravitational Radiation and the Motion of Two Point Masses,''
Ph.D.\ thesis, Caltech, 1964 Feb.\ 14.

\bibitem{Peters:1964zz}
P.~C.~Peters,
``Gravitational Radiation and the Motion of Two Point Masses,''
Phys. Rev. \textbf{136}, B1224-B1232 (1964)
doi:10.1103/PhysRev.136.B1224.

\bibitem{Hansen:1972jt}
R.~O.~Hansen,
``Post-Newtonian Gravitational Radiation from Point Masses in a Hyperbolic Kepler Orbit,''
Phys. Rev. D \textbf{5}, 1021-1023 (1972)
doi:10.1103/PhysRevD.5.1021.

\bibitem{Quinlan:1989}
G.~D.~Quinlan and S.~L.~Shapiro,
``Dynamical Evolution of Dense Clusters of Compact Stars.'' 
Astrophys. J. \textbf{343}, 725-749 (1989).

\bibitem{Mouri2002}
Hideaki Mouri and Yoshiaki Taniguchi, ``Runaway Merging of Black Holes: Analytical Constraint on the Timescale,''
Astrophys. J. \textbf{566}, L17-L20 (2002).

\bibitem{Turner1977}
M.~Turner, ``Gravitational Radiation from Point-Masses in Unbound Orbits---Newtonian Results." 
Astrophys. J. \textbf{216}, 610-619 (1977).

\bibitem{OLeary:2008myb}
R.~M.~O'Leary, B.~Kocsis and A.~Loeb,
``Gravitational Waves from Scattering of Stellar-Mass Black Holes in Galactic Nuclei,''
Mon. Not. Roy. Astron. Soc. \textbf{395}, no.4, 2127-2146 (2009)
doi:10.1111/j.1365-2966.2009.14653.x
[arXiv:0807.2638 [astro-ph]].

\bibitem{Lee1993}
M.~H.~Lee. 
``N-Body Evolution of Dense Clusters of Compact Stars.''
Astrophys, J. \textbf{418}, 147 418 (1993).

\bibitem{Clesse:2016ajp}
S.~Clesse and J.~Garc\'\i{}a-Bellido,
``Detecting the Gravitational Wave Background from Primordial Black Hole Dark Matter,''
Phys. Dark Univ. \textbf{18}, 105-114 (2017)
doi:10.1016/j.dark.2017.10.001
[arXiv:1610.08479 [astro-ph.CO]].

\bibitem{Kovetz:2016kpi}
E.~D.~Kovetz, I.~Cholis, P.~C.~Breysse and M.~Kamionkowski,
``Black Hole Mass Function from Gravitational Wave Measurements,''
Phys. Rev. D \textbf{95}, no.10, 103010 (2017)
doi:10.1103/PhysRevD.95.103010
[arXiv:1611.01157 [astro-ph.CO]].

\bibitem{Ali-Haimoud:2017rtz}
Y.~Ali-Ha\"\i{}moud, E.~D.~Kovetz and M.~Kamionkowski,
Phys. Rev. D \textbf{96}, no.12, 123523 (2017)
doi:10.1103/PhysRevD.96.123523
[arXiv:1709.06576 [astro-ph.CO]].

\bibitem{Gondan:2017wzd}
L.~Gond\'an, B.~Kocsis, P.~Raffai and Z.~Frei,
``Eccentric Black Hole Gravitational-Wave Capture Sources in Galactic Nuclei: Distribution of Binary Parameters,''
Astrophys. J. \textbf{860}, no.1, 5 (2018)
doi:10.3847/1538-4357/aabfee
[arXiv:1711.09989 [astro-ph.HE]].

\bibitem{Rodriguez:2018pss}
C.~L.~Rodriguez, P.~Amaro-Seoane, S.~Chatterjee, K.~Kremer, F.~A.~Rasio, J.~Samsing, C.~S.~Ye and M.~Zevin,
``Post-Newtonian Dynamics in Dense Star Clusters: Formation, Masses, and Merger Rates of Highly-Eccentric Black Hole Binaries,''
Phys. Rev. D \textbf{98}, no.12, 123005 (2018)
doi:10.1103/PhysRevD.98.123005
[arXiv:1811.04926 [astro-ph.HE]].

\bibitem{Raidal:2018bbj}
M.~Raidal, C.~Spethmann, V.~Vaskonen and H.~Veerm\"ae,
``Formation and Evolution of Primordial Black Hole Binaries in the Early Universe,''
JCAP \textbf{02}, 018 (2019)
doi:10.1088/1475-7516/2019/02/018
[arXiv:1812.01930 [astro-ph.CO]].

\bibitem{Sasaki:2018dmp}
M.~Sasaki, T.~Suyama, T.~Tanaka and S.~Yokoyama,
``Primordial Black Holes---Perspectives in Gravitational Wave Astronomy,''
Class. Quant. Grav. \textbf{35}, no.6, 063001 (2018)
doi:10.1088/1361-6382/aaa7b4
[arXiv:1801.05235 [astro-ph.CO]].

\bibitem{Vaskonen:2019jpv}
V.~Vaskonen and H.~Veerm\"ae,
``Lower Bound on the Primordial Black Hole Merger Rate,''
Phys. Rev. D \textbf{101}, no.4, 043015 (2020)
doi:10.1103/PhysRevD.101.043015
[arXiv:1908.09752 [astro-ph.CO]].

\bibitem{Korol:2019jud}
V.~Korol, I.~Mandel, M.~C.~Miller, R.~P.~Church and M.~B.~Davies,
``Merger Rates in Primordial Black Hole Clusters without Initial Binaries,''
Mon. Not. Roy. Astron. Soc. \textbf{496}, no.1, 994-1000 (2020)
doi:10.1093/mnras/staa1644
[arXiv:1911.03483 [astro-ph.HE]].

\bibitem{Jedamzik:2020ypm}
K.~Jedamzik,
``Primordial Black Hole Dark Matter and the LIGO/Virgo observations,''
JCAP \textbf{09}, 022 (2020)
doi:10.1088/1475-7516/2020/09/022
[arXiv:2006.11172 [astro-ph.CO]].

\bibitem{Kritos:2020fjw}
K.~Kritos and I.~Cholis,
``Evaluating the Merger Rate of Binary Black Holes from Direct Captures and Third-Body Soft Interactions Using the Milky Way Globular Clusters,''
Phys. Rev. D \textbf{102}, no.8, 083016 (2020)
doi:10.1103/PhysRevD.102.083016
[arXiv:2007.02968 [astro-ph.GA]].

\bibitem{Kocsis2022}
B.~Kocsis,
``Dynamical Formation of Merging Stellar-Mass Binary Black Holes.'' 
{\em{Handbook of Gravitational Wave Astronomy}} (Springer, Singapore, 2022). pp. 1-44.

\bibitem{Page2024b}
D.~N.~Page,
``Discrete Orbit Effect Lengthens Merger Times for Inspiraling Binary Black Holes,''
in preparation.

\end{thebibliography}
\end{document}